# Bypassing damaged nervous tissue

M.N. Shneider

*Department of Mechanical and Aerospace Engineering, Princeton University, Princeton, NJ, 08544*
e-mail: m.n.shneider@gmail.com

**It is shown the principal possibility of bypassing damaged demyelinated portions of the nervous tissue, thereby restoring its normal function for the passage of action potentials.**

Violations of the nerve fiber integrity, such as partial demyelination and micro-ruptures, leads to malfunction of the nervous system and its components [1-5]. For example, a damage or loss of the myelin coating of neurons, which can be caused by various diseases slows down or even blocks action potentials. This results in a variety of disorders, such as sensory impairment, multiple sclerosis, blurred vision, movement control difficulties, as well as problems with bodily functions and reactions [1-6].

In this letter, it is shown that if a partially demyelinated neuron cell remains alive and functional, appropriate stimulation of the axon away from the damaged area could lead to the normal passage of the action potential through bypassing of the demyelinated area. Such stimulation can be a local change in membrane potential induced by the current in saline, similar to that in neuron synchronization [7]. This can be done, for example, by changing the local probe potential, inducing a current in the saline, resulting in charging of the membrane at a node of Ranvier until the potential difference on the membrane becomes sufficient to initiate an action potential outside of the demyelinated area. Then it becomes possible for further propagation of the action potential along the undamaged segments of the axon. As a result, activity of the nerve will be fully or partly restored.

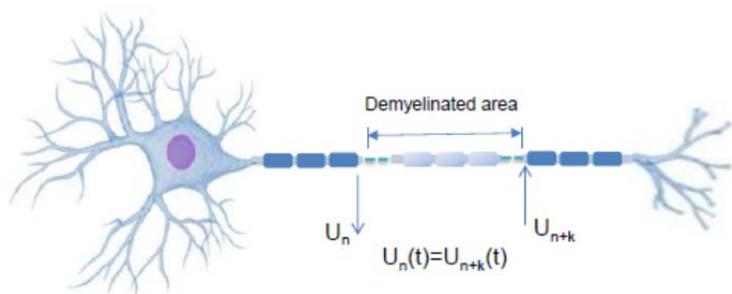

**Fig. 1.** Schematic of a neuron with the characteristic elements of the myelinated axon. The demyelinated portions with damaged segments between *n* and *n+k* nodes of Ranvier are also shown. The arrows indicate the bypass shunting of the demyelinated section.

To examine the action potentials on myelinated nerve fibers, we employ the Goldman–Albus model of a toad neuron [8]. In this model, which has been used by many authors, and reproduced also in [9], a myelinated nerve fiber is represented as a leaky transmission line with uniform internode sections with a length $L_M$ and an outer diameter $d_M$ (myelin sheath inclusive) separated by nodes of Ranvier with a length $L_R$ and outer diameter $d_R$, which is equal to the diameter of an axon inside the myelin coating.

The voltage $U(x,t)$ across the membrane in the internode region is found as a solution to equation [8], which is the equation of potential diffusion:

$$\frac{\partial U}{\partial t} = a \frac{\partial^2 U}{\partial z^2} + bU, \qquad (1)$$

Where $a = 1/R_1 C_1$; $b = -1/R_m C_1$. Here $C_1 = k_1 / \ln(d_M / d_R)$, $R_1 = \frac{R_i}{\pi (d_R/2)^2}$, are the myelin capacitance and resistance per unit length; $R_m = k_2 \ln(d_M / d_R)$ is the myelin resistance times unit length; $k_1, k_2$ are constants and $R_i$ is the axoplasm specific resistance. All parameters and constants are the same as in [8]. The diameters of the myelinated sections of the axon and nodes of Ranvier are $d_M = 15$ and $d_R = 9$ µm, respectively.

The action potential in the nodes of Ranvier $U_{AP}(t)$, with the corresponding resting potential $U_R = -70$ mV, is defined by the Frankenhaeuser–Huxley equation [9,7].

Fig. 2a shows an example of the action potential propagation along an undamaged myelinated axon, calculated in the framework of approximations and data used in [8]. At the model calculations of the damaged axon we assumed for definiteness that demyelination occurs between 12 and 22 nodes of Ranvier. In our calculations we mimic neuron demyelination by assuming a myelin sheath of a smaller thickness. We assume, as an example, that the diameter of myelinated section in demyelinated areas $d_M = 1.025 d_R = 9.225$ µm. That is, in demyelinated areas the capacitance per unit length increases sharply, and the resistance - decreases. Fig. 2b shows the computed results of the action potential in the damaged demyelinated axon. Under these assumed conditions, demyelination causes complete blockage of the action potential.

The conditions for action potential propagation can be restored by shunting of the demyelinated section (nodes 11 and 23 for assumed parameters of our model) (Fig. 2c). In this case, the action potential "skips" the demyelinated area when the maximum potential difference across the membrane reaches ~ -32 mV at the 23$^{rd}$ node. That is, the process develops on the 23$^{rd}$ node of Ranvier as if a pulse with amplitude $U_{cr} \approx 38$ mV and duration of a few milliseconds is applied. This corresponds to approximately 35% of the maximum potential difference across the 11$^{th}$ node of Ranvier membrane. In this case, the action potential propagates down the axon, skipping the demyelinated site.

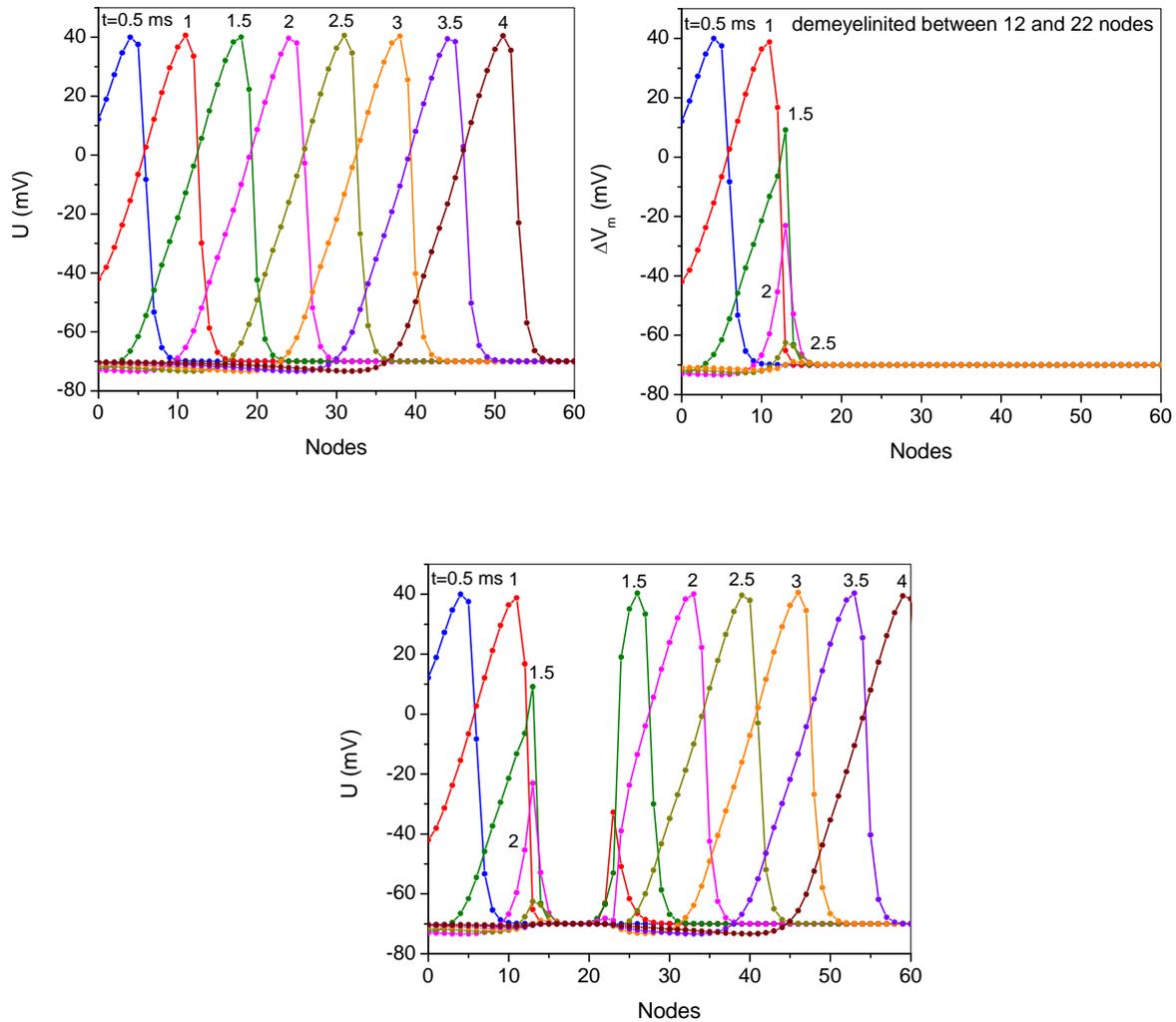

**Fig. 2.** Saltatory propagation of action potential in normal myelin fibers (a); Blocking the action potential in the demyelinated area (b); Bypass of the demyelinated area (between 12 and 22 nodes of Ranvier) (c).

Thus, we have shown that it is possible to bypass damaged demyelinated portions of nervous tissue, thus restoring its normal function. All this is true not only for a single axon, but also for a neuron ensemble, as in the spinal cord, for example.

The results presented in this paper, require experimental verification to evaluate the possibility of its implementation in biomedical practice. To observe the propagation, complete blockage or bypass of the action potential may be possible by detecting the second harmonic generation on a sample of nerve tissue, as suggested in [8,11]. These results demonstrate that the second-harmonic response can serve as a local probe for the state of the myelin sheath, providing a high-contrast detection of neuron demyelination.

It should be noted that it is also possible to block the action potential in an undamaged nerve fiber, leading to reversible anesthesia, while maintaining, at the nodes of Ranvier, a negative voltage below the threshold in some intermediate sections of the axon.